# Pressure-tuned magnetism and conductivity in pyrochlore iridates $Lu_2Ir_2O_7$ and $Er_2Ir_2O_7$

Daniel Staško[*], Petr Proschek, Jiří Prchal, Milan Klicpera

Charles University, Faculty of Mathematics and Physics, Department of Condensed Matter Physics, Ke Karlovu 5, 121 16 Prague 2, Czech Republic

*Corresponding author: daniel.stasko@matfyz.cuni.cz

**Abstract:**

$A_2Ir_2O_7$ iridates were proven to crystallise in the geometrically frustrated pyrochlore structure, which remains stable upon rare-earth cation substitution, temperature variation, and external pressure application. However, the change of interatomic distances and local distortions in the lattice frequently leads to complex electronic properties. The low-temperature behaviour in light-$A$ iridates has been thoroughly investigated, including its evolution with pressure. The present pressure study reports the electrical transport and magnetotransport properties in heavy rare-earth $Lu_2Ir_2O_7$ and $Er_2Ir_2O_7$. Both compounds reveal a semiconductor-to-insulator transition induced by the antiferromagnetic ordering of the all-in-all-out (AIAO) type in the Ir sublattice. The transition monotonously shifts to a higher temperature under applied pressure by approximately 20 K at 3 GPa. As the transition in resistivity originates in the antiferromagnetic order, the latter is expected to be enhanced with the applied pressure as well. Upon cooling the compound in a magnetic field, the AIAO/AOAI domain structure with non-zero net magnetic moment is formed, mirroring itself in an asymmetric term in the magnetoresistance of $Lu_2Ir_2O_7$. The application of pressure then enhances the asymmetric term. The same behaviour is proposed for the whole heavy rare-earth $A_2Ir_2O_7$ series ($A$ = Gd – Lu), although with magnetoresistance features masked significantly by a stronger response of magnetic $A$ cations.

Rare-earth $A_2Ir_2O_7$ iridates crystallising in the geometrically frustrated pyrochlore lattice possess a broad confluence of interactions, i.e., electron correlations, strong spin-orbit coupling (SOC), local Ising anisotropy, or d-f exchange between Ir and $A$ moments. As a result, a plethora of intriguing phenomena, such as Weyl semimetal [1,2], topological Mott insulator [3,4], axion insulator [5], spin-ice state hosting magnetic monopole-like states [6,7], or chiral spin-liquid [8], have been proposed and reported in these materials. The pyrochlore structure (space group $Fd$-$3m$) of the $A_2Ir_2O_7$ iridates with $A$ = Pr-Lu, constituting a ground for mentioned complex electronic states, consists of $A$ and Ir sublattices of interpenetrating corner-sharing tetrahedra and oxygen anions forming the 8- and 6-coordinate cages around respective cations. The cubic lattice parameter $a$ and the single fractional coordinate of the oxygen $x_{48f}$ stand for the only free parameters of the structure. The stability of the pyrochlore structure has been demonstrated throughout the whole rare-earth series, down to low temperatures and at pressures up to 20 GPa [9,10,11,12]. Except for the $A$ = Pr member revealing no magnetic transition [1,13], $Ir^{4+}$ sublattice orders antiferromagnetically (AFM) with the so-called all-in-all-out (AIAO) magnetic ordering in the $A_2Ir_2O_7$ iridates [10,14,15]. The transition temperature $T_{Ir}$ increases monotonously with the atomic number of the rare earth from approximately 39 K ($A$ = Nd) to 147 K ($A$ = Lu), depending rather on the radius of $A$, and related change of $a$ and $x_{48f}$ parameters, than on the magnetic characteristics of the rare-earth cation [13,16]. The AIAO ordering induces a concomitant change in the material's conductive properties. The metal-to-insulator transition has been observed in light rare-earth members [13,17,18]. In heavy $A_2Ir_2O_7$ with a larger conduction gap, the transition from a nonmetallic or semiconducting state to an insulating state has been reported [13,14,19]. The metal-to-insulator transition was recently ascribed to a gap opening at $T_{Ir}$ (or slightly lower temperature, marked as $T_{MI}$ in our previous study) via the Slater mechanism [19,20,21]. Still, the low-temperature state in $A_2Ir_2O_7$ cannot be characterised as a typical Slater insulator [19,20]; a crossover from the Slater to Mott regimen is anticipated upon further cooling the system below $T_{MI}$ [21].

Complex and conjoined magnetic and conducting properties of the compound can be conveniently investigated using the magnetoresistance technique, as illustrated by a number of $A_2Ir_2O_7$ studies [1,6,18,22,23,24,25]. In addition to the above-discussed properties, the Ir sublattice's AIAO/AOAI domain structure and its interactions with the magnetic $A^{3+}$ cations on the rare-earth sublattice have been revealed. Notably, an asymmetric magnetoresistance has been observed in $Eu_2Ir_2O_7$ after cooling the sample below $T_{Ir}$ in an applied magnetic field [22,23,24,25], evidencing a robust induced ferromagnetic component on the AIAO/AOAI interface. Magnetoresistance of the $Ho_2Ir_2O_7$ member has revealed a strong interaction between AIAO/AOAI domains and interfaces and Ho magnetic sublattice with proposed magnetic monopole-like states [6]. The quantum critical point induced by applied pressure and related phase diagram in $Sm_2Ir_2O_7$, undetected in zero-field transport properties, have been mapped using magnetoresistance measurements [17].

The application of external pressure on the system represents a way to manipulate its electronic properties, in addition to more standard variables like temperature and external magnetic field. Generally, the pressure suppresses the magnetism in the compound, possibly resulting in a quantum critical point, as observed in $Sm_2Ir_2O_7$ [11,17]. The compression of the lattice in applied pressure enhances the 5d bandwidth of iridium, and thus the hopping parameter; that is, the insulating behaviour of $A_2Ir_2O_7$ turns more metallic at higher pressures (experimentally proved in $A$ = Nd, Sm and Eu) [17,26,27]. Moving from the light to the heavy rare-earth members, the effect of pressure on the electronic properties becomes more complex. The metal-to-insulator transition temperature $T_{MI}$ decreases with the applied pressure in $A$ = Sm, stagnates in $A$ = Eu, and increases in $A$ = Gd member [28]. Simultaneously, $T_{Ir}$ increases slightly with pressure up to 2 GPa and decreases again at a higher pressure in $Eu_2Ir_2O_7$ [29], indicating competing contributions stabilising the AFM state. The electronic or magnetic properties of the heavy-rare-earth members $A$ = Tb - Lu have not been investigated in applied pressure so far.

The present study reports the transport and magnetotransport properties of the heavy-rare-earth $A_2Ir_2O_7$ ($A$ = Er and Lu) at hydrostatic pressure of up to 3 GPa. From the viewpoint of the pyrochlore crystal structure, the heavy-rare-earth analogues have the most distorted local structure (interatomic distances and bond angles connected with parameters $a$ and $x_{48f}$) [9,14,16]. In turn, they exhibit the highest $T_{Ir}$. The $Lu_2Ir_2O_7$ end-member was selected for its electronic and magnetic properties stemming from the $Ir^{4+}$ sublattice only, similar to the light $A$ = Eu member. $Er_2Ir_2O_7$ hosts, in addition, a substantial magnetic moment of $Er^{3+}$, which dictates the compound's magnetism, especially at low temperatures. Investigating the two compounds, therefore, allows us to encompass the conducting and magnetic properties of the heavy rare-earth part of the $A_2Ir_2O_7$ in general. The present results complete the picture of magnetotransport properties in the $A_2Ir_2O_7$ series, their systematics and variability with external pressure.

Compared to the light-rare-earth members with a clear metal-to-insulator anomaly in the temperature evolution of the electrical resistivity [13,17,18], heavy $A_2Ir_2O_7$ reveal only a relatively broad feature in resistivity data observable in a log-log scale (Fig. 1a-b). The transition from a semiconducting state to an insulating state manifests as a change of slope on the resistivity curve. $Lu_2Ir_2O_7$ exhibits the broadest anomaly and simultaneously the most insulating behaviour out of the rare-earth series. The change of resistivity of 6-7 orders of magnitude between room temperature and low temperatures is revealed (Fig. 1a, see also Supplementary materials for details on resistivity measurements, as well as sample synthesis and characterisation). In contrast, only about three orders of magnitude change is followed in $Er_2Ir_2O_7$. The application of external pressure monotonously reduces the electrical resistivity of both compounds; the reduction is more pronounced at low temperatures, especially in $Lu_2Ir_2O_7$. As pelletised powder samples were measured, the reduction of resistivity is partially explained by the grain structure; that is, pressure improves conductivity on the grain boundaries as the powder is compressed into a more compact state at higher pressure. Simultaneously, the compression of the lattice contributes to a more conducting state, which is portrayed in the low-temperature region (below ~10 K), where pressure changes resistivity by 1-2 orders of magnitude. Nevertheless, 3 GPa is clearly insufficient pressure to turn an insulating state metallic.

Increasing the atomic number of $A$ results in a broad semiconductor-to-insulator anomaly in $A_2Ir_2O_7$ resistivity; therefore, reasonably determining the transition temperature $T_{MI}$ becomes tricky. Determining $T_{MI}$ from the derivative $d\ln(\rho)/d(T^{-1/4})$, serving its purpose well in $A$ = Sm, Eu, Gd members [28], is unreliable in heavy-$A$ analogues. As the electrical resistivity in the log-log scale reveals mostly linear development at temperatures above and below the anomaly, simple linear fits of these regions and the intersection of extrapolated lines have been used to estimate $T_{MI}$ [19,30]. Although such a determination of $T_{MI}$ cannot be physically justified, it allows us to follow the shift of the anomaly with the applied pressure. Clearly, $T_{MI}$ is monotonously increasing with the application of pressure in both $Lu_2Ir_2O_7$ and $Er_2Ir_2O_7$ (Fig. 1c). The same evolution was reported for the $A$ = Gd member [28]. In contrast, the light $A$ = Sm member reveals $T_{MI}$ decreasing with increasing pressure. $Eu_2Ir_2O_7$ stands on a border between the two regimens, showing almost no evolution of $T_{MI}$ with external pressure [28]. As the semiconductor-to-insulator transition at $T_{MI}$ is tied to the AFM ordering below $T_{Ir}$, we anticipate that the application of pressure strengthens the AFM order ($T_{Ir}$ increases) in all heavy rare-earth $A$ = Gd – Lu pyrochlore iridates. Notably, $T_{MI}$ in $Lu_2Ir_2O_7$ is determined to have a lower value than in $Er_2Ir_2O_7$ in all pressures, which we ascribe to the broadness of the anomaly and the way the transition temperature is estimated. $T_{MI}$ temperatures were estimated to be approximately the same in other $A_2Ir_2O_7$ [30], measuring the conductive properties of these compounds in ambient conditions. $T_{Ir}$, determined from the magnetisation data measured on the same polycrystalline samples, increased monotonically with $A$ [16,19].

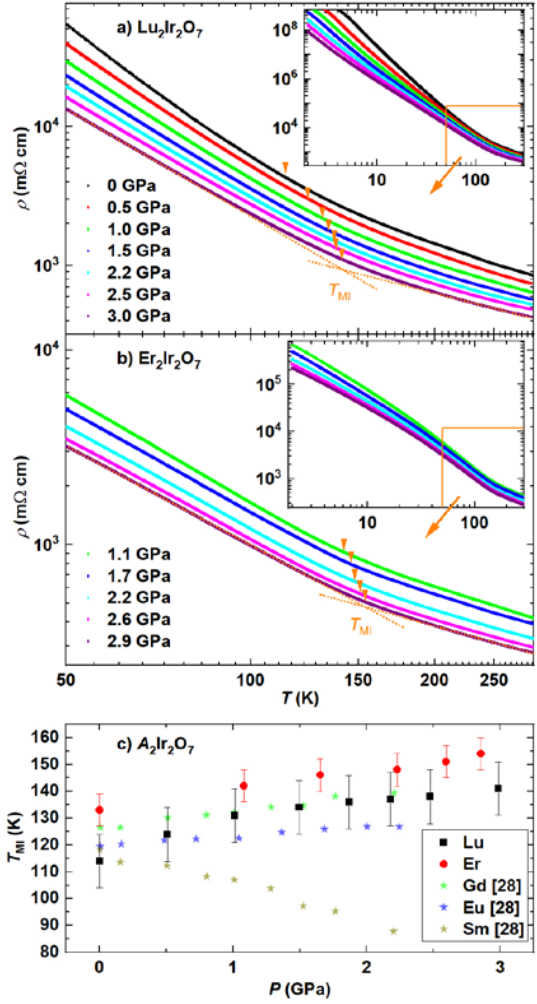

Fig. 1: Temperature dependence of electrical resistivity of a) $Lu_2Ir_2O_7$ and b) $Er_2Ir_2O_7$ under pressure. The whole temperature region and zoomed-on transitions are presented, respectively. The temperature of the transition $T_{MI}$ (indicated by arrows) was estimated by linear fitting of the high- and low-temperature regions (indicated by dotted lines). c) $T_{MI}$ increases with the applied pressure. The pressure evolutions of $T_{MI}$ in light-rare-earth $A$ = Sm, Eu and Gd members [28] are included as well.

The effects of external and chemical pressure on the transition temperature $T_{MI}$ in heavy $A_2Ir_2O_7$ were revealed to be quite distinct. Substituting Er for Yb in $Er_2Ir_2O_7$ results in a contraction of the unit cell similar to the application of 3 GPa of pressure [9]. However, the substitution was shown to have only a moderate impact on the transition temperature $T_{Ir}$ ($T_{MI}$) [16,19,31,32], while 3 GPa pressure shifts $T_{MI}$ to higher temperatures by more than 20 K. The inequality of external and chemical pressure is demonstrated, but the lattice parameter $a$ plays only a minor role in determining $T_{MI}$. The AFM ordering of the Ir sublattice stems mainly from the electron correlations $U/t$ and the local Ising spin anisotropy [11]. A decrease of the lattice parameter $a$ increases the hopping integral $t$ and, therefore, suppresses the AFM ordering. In turn, the semiconductor-to-insulator transition, directly connected with magnetic ordering, shall also shift to a lower temperature. Such behaviour might be supported by slight/no change of $T_{MI}$ with $A$ substitution in heavy

$A_2Ir_2O_7$ [19]. In contrast, the previous magnetisation study revealed a small increase of $T_{Ir}$ with an increasing atomic number of $A$ [16]. The applied pressure, on the contrary, enhances the transition. Therefore, the second structural parameter, $x_{48f}$, dictating the distortion of the oxygen cages around $Ir^{4+}$ cations, is responsible for observed evolutions. The impact of the oxygen cage distortion on the conducting and magnetic properties of pyrochlore iridates was repeatedly discussed [9,17,28]. The parameter $x_{48f}$ is proposed to play a crucial role in strengthening the local Ising anisotropy in the system. $x_{48f}$ slightly increases with Er by Yb substitution [9,16], while almost no change of the parameter with applied pressure has been reported [9,11,33]. That is, the oxygen cages around iridium cations are less distorted by applied pressure than by substitutions. It enhances the local Ising anisotropy, ordering temperature $T_{Ir}$ and, in turn, $T_{MI}$. The same behaviour was observed in several $A_2Ir_2O_7$ members, and we propose it to be general in the whole heavy-rare-earth $A$ = Eu – Lu series. The opposite evolution of $T_{MI}$ with pressure in light-rare-earth $Sm_2Ir_2O_7$ [28] could be understood considering its larger unit cell and a larger Ir-O-Ir bond angle.

In addition to conducting properties, the magnetic properties of $Lu_2Ir_2O_7$ and $Er_2Ir_2O_7$ were investigated using the magnetotransport measurement technique. The magnetic field was applied parallel to the electrical current, minimising the Hall contribution to the signal. The formation of the AFM order below $T_{Ir}$ was observed to influence the magnetoresistance significantly. Employing the three cooling protocols ZFC, 9 T FC, and -9 T FC (see details in Supplementary materials), a considerably different magnetoresistance was followed in $Lu_2Ir_2O_7$ (Fig. 2). Cooling the sample in a zero field through $T_{Ir}$, the magnetoresistance has a symmetrical shape. A similar evolution in the ZFC regimen was reported for other non-magnetic $A$ members of the series, i.e., $Eu_2Ir_2O_7$ (comparable only at higher temperatures) [22,23] and $Y_2Ir_2O_7$ [2]. Cooling the sample through $T_{Ir}$ in an applied field (FC regimens) results in an asymmetric magnetoresistance response dependent on the strength and sign of the external field. The asymmetry is proposed to originate from an additional ferromagnetic contribution of the AIAO/AOAI domain interfaces [22,34] induced by the FC protocol. Such an explanation is supported by the fact that the asymmetric FC magnetoresistance becomes almost symmetric (similar to ZFC magnetoresistance) when a linear contribution (dashed lines in Fig. 2) is subtracted. Importantly, the ferromagnetic component on the interface is robust against the considerably high magnetic field, being protected by the antiferromagnetic AIAO and AOAI domains [34]. Indeed, a linear term in magnetoresistance would, otherwise, be unstable in high magnetic fields; the magnetoresistance would become symmetric.

In addition to the asymmetric FC contribution present at temperatures below $T_{Ir}$ ($T_{MI}$), a hysteresis in magnetoresistance is observed (Fig. 2). While there is no clear hysteresis at the lowest temperatures, at intermediately high temperatures (50 K data presented in Fig. 2b) a clear difference between data measured with increasing and decreasing magnetic field is revealed. A similar behaviour, both the symmetric and asymmetric magnetoresistance and the hysteresis at higher temperatures, were previously observed in $Eu_2Ir_2O_7$ single crystals [22] and thin films [23,24]. Hysteresis in magnetoresistance can be generally explained by a finite coercive field created by intrinsic ferromagnetism; in $A_2Ir_2O_7$, it is connected with the AFM domains and their ferromagnetic interfaces (the domain wall model [18,34,35]). However, the net FM moment responsible for the linear contribution discussed above does not induce the hysteretic behaviour, as the hysteresis is also observed in the ZFC data at 50 K. Instead, rotatable FM moments at the domain boundaries are considered. The robust net magnetic moment connected with a three-in-one-out (3I1O) or a two-in-two-out (2I2O) ordering of magnetic moments in the interface is accompanied by moments at the boundaries weakly coupled with domains, which an applied magnetic field can easily rotate [22]. An analogous scenario was proposed to explain the EB effect in AFM/FM heterostructure interfaces. There, the asymmetric term and hysteresis are attributed to the pinned and rotatable FM contributions, respectively [36,37]. The absence of hysteresis in our 2 K data could then be attributed to a freezing of the respective rotatable moments, in agreement with the increase of the asymmetric linear term upon cooling.

Alternatively, the linear asymmetric term was previously ascribed solely to the AIAO/AIAO domain structure and explained by the double exchange model [23,38], although it cannot explain the observed hysteresis. Notably, the asymmetry and hysteresis in magnetoresistance seem analogous to the magnetisation response in $A$ = Eu, Er and Lu members [22,34,39], where a robust shift and a hysteresis in magnetisation is dependent on the strength and direction of the FC cooling field.

Low-temperature magnetoresistance in $Er_2Ir_2O_7$ is mostly dictated by the $Er^{3+}$ magnetism, resulting in approximately a 30-40% resistivity decrease upon applying a 9 T field at 2 K (Fig. S1 in Supplementary materials). A large resistivity drop in the applied field (in contrast with a few percent drop in common metals [40]) is typical for the $A_2Ir_2O_7$ iridates with magnetic rare-earth elements; e.g., more than 90% drop in resistivity was reported in $Dy_2Ir_2O_7$ [19], or a giant magnetoresistance present in $Nd_2Ir_2O_7$ [41]. No clear ferromagnetic contribution of the iridium sublattice AIAO/AOAI interfaces was observed in magnetotransport properties, being most likely masked by a significantly higher Er contribution. Applying external pressure, the magnetoresistance value first decreases and above 1.7 GPa again slightly increases. Only a decreasing tendency with increasing pressure was observed in $Lu_2Ir_2O_7$, pronouncing the effect of the magnetic Er sublattice (pressure-affected local environment of cations) on the magnetoresistivity of the compound.

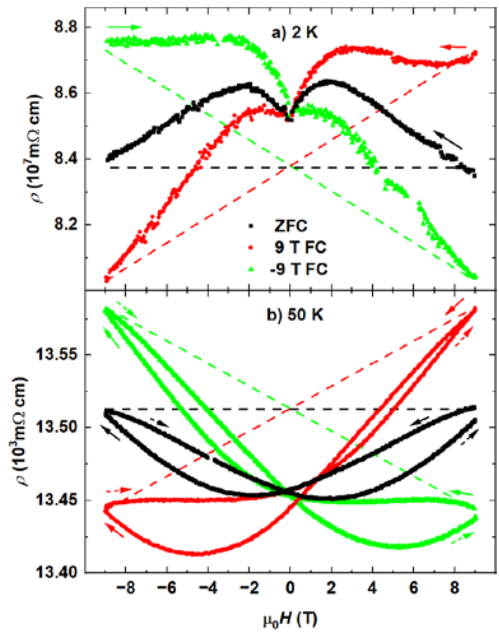

Fig. 2: Magnetoresistance measured in $Lu_2Ir_2O_7$ at a) 2 K and b) 50 K under 3 GPa of pressure. The sample was cooled in fields of 0 T (ZCF regimen), 9 T (9 T FC regimen), and -9 T (-9 T FC regimen) from 300 K down to 2 K (see details in Supplementary materials). ZFC magnetoresistance is symmetric in a magnetic field; FC magnetoresistances show an additional linear contribution, which is positive or negative, depending on the cooling regimen (depicted by the dashed lines). No clear hysteresis is observed at 2 K, while a significant hysteresis is followed at 50 K.

The pressure effect on the magnetoresistance in $Lu_2Ir_2O_7$ is documented by a monotonous change of the asymmetric linear term (Fig. 3). Note that the absolute value of the resistance changes significantly with the applied pressure (Fig. 1); therefore, the relative magnetoresistance is presented in Fig. 3a. Although the shape of the magnetoresistance curve remains similar up to 3 GPa, overall changes in magnetoresistance are enhanced with pressure. To quantify the evolution of the asymmetric part, a linear coefficient $\alpha$ was determined, employing the equation $\rho(\mu_0H)/\rho(0) =$ (symmetric part) $+ \alpha\mu_0H$ [23,38]. A visual representation of the linear part is depicted as dashed lines in Fig. 2. The coefficient $\alpha$ increases strongly with the applied pressure at 2 K. At 50 K, $\alpha$ has an order of magnitude lower value (Fig. 3b-c). Extrapolating $\alpha$ down to ambient pressure, similar values were previously reported in nonmagnetic-rare-earth $Eu_2Ir_2O_7$ [23]. The increase of the $\alpha$ value with pressure is ascribed to the enhanced Ir domain structure with ferromagnetic domain interfaces under applied pressure [23,25,38]. An increase in asymmetry with pressure is thus indirectly connected with the $T_{MI}$ increase (Fig. 1). Lower asymmetry at higher temperatures is tentatively ascribed to the presence of the rotatable magnetic moments on the domains' boundaries, responsible for the hysteresis in magnetoresistance as discussed above.

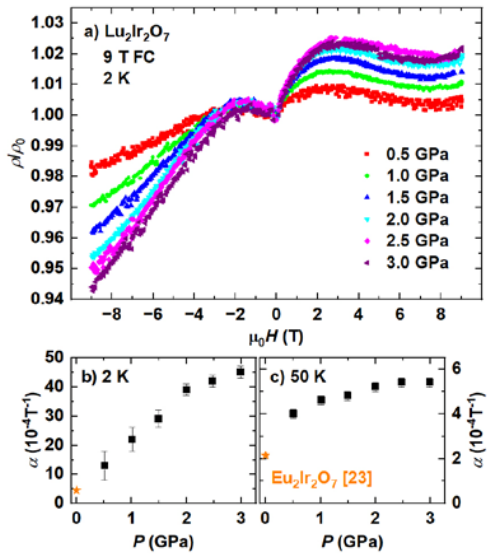

Fig. 3: Relative magnetoresistance of $Lu_2Ir_2O_7$ measured using the 9 T FC protocol at various external pressures. a) 2 K magnetoresistance data are plotted. The application of pressure enhances the linear asymmetric term (see also Fig. 2), which is described by the parameter $\alpha$ presented in panels b)-c) for measurements at 2 K and 50 K. Data from $Eu_2Ir_2O_7$ thin films at ambient pressure [23] are included for comparison.

To summarise, we investigated the pressure evolution of the transport and magnetotransport properties of two polycrystalline heavy-rare-earth $Lu_2Ir_2O_7$ and $Er_2Ir_2O_7$ iridates to complement previous studies on light $A_2Ir_2O_7$. Between the room temperature and 2 K, the electrical resistivity increases by 6-7 orders of magnitude in the $A$ = Lu (highest increase among the $A_2Ir_2O_7$ compounds) and by three orders of magnitude in the $A$ = Er member. Both compounds exhibit a broad semiconductor-to-insulator transition connected to the AFM ordering of the Ir sublattice. The temperature of the transition increases with pressure, similarly to the case of $Gd_2Ir_2O_7$, setting the trend for all heavy-rare-earth iridates. The pressure evolution of the transition temperature is discussed within the framework of the local structural environment in the compound, most importantly the local distortion of the $IrO_6$ cages, and compared to the distinct evolution with chemical pressure. Magnetoresistance measurements revealed the asymmetric feature and hysteresis connected to the AIAO/AOAI domain structure in $Lu_2Ir_2O_7$, being general in the whole $A_2Ir_2O_7$ series (previously observed in $Eu_2Ir_2O_7$). In $A_2Ir_2O_7$ with magnetic $A$ cation, these effects are most likely present but masked by stronger rare-earth magnetism. The pressure evolution of the asymmetric term in magnetoresistance is in good agreement with the pressure-enhanced AFM state.

**Data availability:**

The raw/processed data required to reproduce these findings are available upon reasonable request.

**Acknowledgements:**

Sample synthesis and characterisation were done in MGML (mgml.eu), which is supported within the program of Czech Research Infrastructures (project no. LM2023065). The study was supported by Charles University, project GA UK (no. 148622), and the Barrande Mobility project (no. 8J24FR013).

# SUPPLEMENTARY MATERIALS

## Pressure-tuned magnetism and conductivity in pyrochlore iridates $Lu_2Ir_2O_7$ and $Er_2Ir_2O_7$

Daniel Staško[*], Petr Proschek, Jiří Prchal, Milan Klicpera

Charles University, Faculty of Mathematics and Physics, Department of Condensed Matter Physics, Ke Karlovu 5, 121 16 Prague 2, Czech Republic

*Corresponding author: daniel.stasko@matfyz.cuni.cz

**Sample synthesis and characterisation:**

Polycrystalline samples of $Lu_2Ir_2O_7$ and $Er_2Ir_2O_7$ were synthesised by the CsCl flux method from a stoichiometric mixture of $A_2O_3$ rare-earth oxides and $IrO_2$ (99.99% metal basis, Alfa Aesar). Details on the synthesis were published in our recent paper [42]. The same samples (powder from the same batch) were studied in our very recent publications [19] (transport properties at ambient pressure) and [9] (structural properties at extreme pressure/temperatures), as well as in publications [16,43]. Sample stoichiometry, crystallinity and phase purity were checked by employing a scanning electron microscope (TESCAN, Mira I LMH) with an energy-dispersive X-ray (EDX) detector. The ratio $A$:Ir = 50(1):50(1) was confirmed. The oxygen content cannot be quantitatively evaluated using the EDX method. The pyrochlore crystal structure (space group *Fd-3m*) was verified by X-ray powder diffraction employing a Bruker diffractometer with Cu K$\alpha$ radiation. A small amount of minority $A_2O_3$, $IrO_2$ and Ir phases were observed in the diffraction patterns, being commonly present in $A_2Ir_2O_7$ materials [10,29,39,42]. The structure was investigated further by employing synchrotron radiation facilities, determining ambient lattice parameter a = 10.10154(2) Å and oxygen fractional coordinate $x_{48f}$ = 0.3395(3) for $Lu_2Ir_2O_7$, and a = 10.16556(23) Å and $x_{48f}$ = 0.338(3) for $Er_2Ir_2O_7$ [9].

**High-pressure experimental methods:**

The investigated pellets were prepared by cold pressing the powder with approximately 2 kbar of pressure. Prism-shaped samples of approximate length of 3 mm and a cross-section of 1x1 mm$^2$ were cut from the pellets. The four-probe method was employed to measure electrical resistivity, with approximately 1 mm distance between voltage contacts. Copper wires were attached to the samples using a Dupont silver conducting paste. A hybrid piston-cylinder pressure cell was used for the application of hydrostatic pressure up to 3 GPa. The sample was inserted in the pressure cell together with a manganin manometer and submerged in the Daphne 7575 oil [44], which acted as the pressure-transmitting medium. The pressure medium remains liquid at room temperature under pressure up to 3.9 GPa, ensuring the application of pressure happens in hydrostatic conditions. The measurements were done employing the commercial PPMS (Quantum design), able to reach temperatures down to 2 K and magnetic field up to 9 T. Electrical resistivity was measured using the Delta method via Keithley 6221 (current source) and Keithley 2182A (nanovoltmeter). Complimentary measurements were performed using a closed-cycle refrigerator (CCR; Sumitomo Heavy Industries / Janis Research). The CCR is optimised for the resistivity measurements in high-pressure cells but lacks the possibility to apply a magnetic field. Qualitatively the same data compared to the ones collected using the PPMS were measured.

A freshly prepared sample was used in the case of $Er_2Ir_2O_7$, while the same sample from an ambient pressure measurement at low temperature [19] was reused in the case of $Lu_2Ir_2O_7$. Notably, the integrity of the sample surface and the contacts was more robust for the reused sample compared to the fresh one. Small anomalies (smoothed for clarity in Fig. 1) attributed to the solidification and melting of the Daphne 7575 oil were observed in the $Er_2Ir_2O_7$ data, in perfect agreement with the phase diagram of the Daphne 7575 oil [44]. In comparison, the $Lu_2Ir_2O_7$ showed smoother behaviour without anomalies, supposedly thanks to being cycled down to 2 K before the contact with the pressure-transmitting medium.

The electric current was applied along the magnetic field direction to suppress the Hall contribution. At low temperatures and ambient(low) pressure, the electrical resistivity was too large to be reliably determined in $Lu_2Ir_2O_7$. That is, the experimental method and the apparatus were reaching the limit when an electrical current in orders of nA was seemingly enough to locally heat up the sample in the lowest temperature region. Therefore, data from only a reasonable resistivity range are shown in the inset of Fig. 1a.

Magnetoresistivity was measured on samples cooled using three protocols: a zero-field-cooled regimen (ZFC) and two field-cooled regiments (9 T FC and -9 T FC). The sample was cooled from 300 K down to the desired temperature in zero magnetic field (ZFC), a field of 9 T (9 T FC), or a field of -9 T (-9 T FC). The ZFC sample was measured in a magnetic field $0 \rightarrow 9 \rightarrow -9 \rightarrow 0$ T at 2 K, heated to 50 K, and measured in a field $0 \rightarrow 9 \rightarrow -9 \rightarrow 9$ T. The 9 T FC sample was measured in a magnetic field $9 \rightarrow -9 \rightarrow 0$ T at 2 K, heated to 50 K, and measured in a field $0 \rightarrow 9 \rightarrow -9 \rightarrow 9$ T at 50 K. The -9 T FC was measured in a magnetic field $-9 \rightarrow 9 \rightarrow 0$ T at 2 K, heated to 50 K, and measured in a field $0 \rightarrow 9 \rightarrow -9 \rightarrow 9$ T at 50 K. Arrows mark the sense of magnetic field variation during the measurement in Fig. 2.

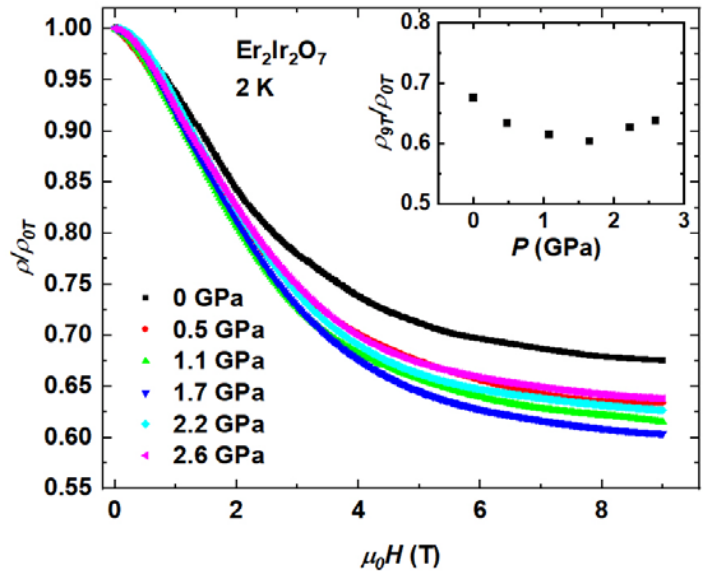

Fig. S1: Magnetoresistance of $Er_2Ir_2O_7$ at 2 K under the applied external pressure. Magnetoresistance tends to saturate approaching the field of 9 T. The observed 30-40% drop in resistivity upon the application of the 9 T field is similar to the drop in sister $A_2Ir_2O_7$ iridates with magnetic rare-earth elements [19].